\documentclass[12pt]{article}
\usepackage{graphicx}
\usepackage[bookmarksnumbered,colorlinks,plainpages]{hyperref}

\begin{document}

\title{On different $q$-systems in nonextensive thermostatistics}

\author{W. Li$^{\dag\ddag}$, Q.A. Wang$^{\dag}$,
L. Nivanen$^{\dag}$, and A. Le M\'ehaut\'e$^{\dag}$\\
{\small $^{\dag}$Institut Sup\'erieur des Mat\'eriaux du Mans}, \\
{\small 44, Avenue F.A. Bartholdi, 72000 Le Mans, France}\\
{\small $^{\ddag}$Institute of Particle Physics}, \\
{\small Hua-Zhong Normal University, Wuhan 430079, P.R. China}}

\date{}

\maketitle

\begin{abstract}
It is known that the nonextensive statistics was originally formulated for the
systems composed of subsystems having same $q$. In this paper, the existence of
composite system with different $q$ subsystems is investigated by fitting the power
law degree distribution of air networks with $q$-exponential distribution. Then a
possible extension the nonextensive statistics to different $q$ systems is provided
on the basis of an entropy nonadditivity rule and an unnormalized expectation of
energy.
\end{abstract}


\section{Introduction}
The starting point of the nonextensive statistics (NS) of Tsallis\cite{Tsal88} is
the entropy given by
\begin{equation}                                    \label{1}
S_q=\frac{\sum_{i=1}^wp_i^q-1}{1-q}
\end{equation}
where the physical states are labelled by $i=1,2,...,w$, $q$ is a parameter
characterizing the nonextensivity of the theory, and $p_i$ is the probability for
the system to be found at state $i$. The application of the principle of maximum
entropy under appropriate constraints can lead to power law distributions
characterized by the so called $q$-exponential functionals\cite{Tsal88,Tsal98}. For
example, for canonical ensemble under the constraints associated with probability
normalization and average energy, we can have :
\begin{equation}                                    \label{4}
p_i=\frac{1}{Z_q}[1-(1-q)\beta E_i]^{1/1-q} \;\;\; [\cdot]>0
\end{equation}
where $Z_q$ is the partition function, $E_i$ is the energy of a microstate $i$, and
$\beta=\frac{\partial S_q}{\partial U_q}$ is the Lagrange multiplier associated with
the average energy $U_q$. It is worth noticing that, due to the different formalisms
of NS proposed in the past 15 years, the inverse temperature $\beta$ has been given
several definitions in different formulation of NS. A comment on this subject was
given in ref.\cite{Wang04a}. In the present work, we consider the formulation with
$\sum_{i=1}^wp_i=1$ and $U_q=\sum_{i=1}^wp_i^qE_i$ leading to the $q$-exponential
distribution given by Eq.(\ref{4}).

For a composite system $(A+B)$ whose joint probability $p_i(A+B)$ is given by the
product of the probabilities of its subsystems {\it having the same $q$}, i.e.,
$p_{ij}(A+B)=p_i(A)p_j(B)$ (product joint probability-PJP), we have, from
Eq.(\ref{1})
\begin{equation}                                    \label{2}
S_q(A+B)=S_q(A)+S_q(B)+(1-q)S_q(A)S_q(B)
\end{equation}
and, from Eq.(\ref{4})
\begin{equation}                                    \label{3}
E_{ij}(A+B)=E_i(A)+E_j(B)+(q-1)\beta E_i(A)E_j(B)
\end{equation}
Eqs.(\ref{2}) and (\ref{3}) prescribe the class of nonextensive systems to which NS
may be applied. They are analogs (or generalizations) of the entropy and energy
additivity in the conventional Boltzmann-Gibbs statistics (BGS) and can be
considered as starting hypothesis of NS. It has been proved that\cite{Santos,Abe00x}
Eq.(\ref{2}) uniquely determines Tsallis entropy (given PJP) and
that\cite{Abe02a,Wang02a} Eqs.(\ref{2}) and (\ref{3}) are a group of necessary
conditions for the existence of thermal equilibrium and stationarity between
nonextensive systems.

Among the fundamental questions about NS, an interesting one is about its validity
for the systems containing subsystems with different $q$'s, as has been recently
discussed by several authors\cite{Tsal03,Sasaki,Nauenberg}. The debate turns around
the establishment of the zeroth law for nonextensive systems. As is well known, for
equilibrium (and local equilibrium) systems, this law allows one to measure
intensive variables like temperature and pressure and to relate them to other
thermodynamic variables and functions. For other nonequilibrium systems, a
relationship between intensive variables like zeroth law may also be necessary in
order to maintain the links between different parts of a system through
thermodynamic functions.

The applications of NS during the last years were often carried out for systems
considered as a whole\cite{Beck2,Plastinos,Lavagno,Kaniadakis}. So the question of
composition of different $q$ subsystems did not arise. From theoretical point of
view, hierarchically invariant NS, just like BGS, should have its place for the
hierarchies that have about the same space-time scale and include essentially the
same physics. An example is the networks to which application of NS has been
considered\cite{Tsallis04}. Indeed, in the study the scientific collaboration
networks\cite{Barabasi} for example, if NS is successful for an international
collaboration network, it is hoped that it holds also for the national or regional
collaboration networks which compose the international one and possibly have
different $q$'s. The reader will find below an example of this situation with the
airline networks.

A possible extension of NS was briefly mentioned in \cite{Wang04a}. In what follows,
after demonstration of the existence of different $q$ subsystems and composite systems
all obeying $q$-exponential distribution given by Eq.(\ref{4}), a detailed development
of this extension is given by replacing Eq.(\ref{2}) with a new nonadditivity rule and
by establishing the zeroth law with the unnormalized expectation of energy as
mentioned above. It is expected that this result may serves as a possible mathematical
complement of the work of \cite{Tsal03,Moyano} where the authors studied the
possibility of measuring the dynamical temperature of a non Boltzmannian systems
($q\neq 1$) in stationary states with a Boltzmannian thermometer ($q=1$). This work is
not complete as a generalization of the whole NS theory to different $q$ systems
because the validity of this approach for the normalized expectation based on escort
probability\cite{Tsal98}, widely usd in NS, is still open for investigation.

\section{An example of different $q$-systems}
Here we present an application of NS to both a total system and its subsystems, in
order to show that the two subsystems and the composite system all have
$q$-exponential distribution but with different $q$'s. Two subsystems are China air
network and US air network. The composite system is the air network which includes
the airports in both countries.

In the air network, airports are viewed as nodes. The connections between airports
are simply represented by flights. In the terminology of network, the degree $k$ of
a certain airport means it has flights with $k$ other airports in the same network.
A very important quantity is the distribution of $k$, $p(k)$, called degree
distribution, which tells the probability of finding an airport with degree $k$.
Apparently, the degree distribution specifies the topology of the network
investigated. For example, there are many different types of networks, classified by
their respective degree distributions, scale-free, random, and so on.

China air network consists of 128 major airports \cite{cn-airnetwork} and for US air
network, that number is 215 \cite{us-airnetwork}. Hence the composite system of
China and US air network has 343 major airports. Besides all the flights of the two
original subsystems, the newly composed system also includes a few international
flights, such as the one from Beijing to New York, etc. Since the number of
international flights is much less than that of national ones, constructing the
composite system can be viewed as adding two independent subsystems.

Two main reasons may account for why we can use the Tsallis distribution to fit the
degree distribution. First, the air network is not a system which can reach the
state of equilibrium. Like many other complex systems in evolution, the air network
consists of many units having complicated interplay (interactions). Second, the
observations from \cite{cn-airnetwork} and \cite{us-airnetwork} have suggested that
the degree distributions behave more like power-laws (in two regimes). As we know,
Tsallis statistics provides a way to generate naturally power-law distribution and
has been satisfactorily used for fitting certain distribution functions of complex
systems\cite{TsallisGeorgeMendes,Borges}.

Our first idea was to fit the observed degree distributions separately in the two
regimes with the following function
\begin{equation}
p(k)=\frac{[1-(1-q_i)\beta_i k]^{\frac {1}
{1-q_i}}}{\sum_{k}[1-(1-q_i)\beta_i k]^{\frac {1} {1-q_i}}}, i=1,2
\end{equation}

\noindent where $q_1,\beta_1$ and $q_2, \beta_2$ are the parameters for the small
$k$ and large $k$ regimes, respectively. $(\beta_1, \beta_2)$ are (0.46$\pm$0.005,
2.85$\pm$0.01), (0.67$\pm$0.003, 3.34$\pm$0.02) and (0.61$\pm$0.003, 4.05$\pm$0.02),
for China air network, US air network and the composite air network, respectively.
The corresponding values of $(q_1, q_2)$ are (3.16$\pm$0.01, 1.35$\pm$0.007),
(2.49$\pm$0.01, 1.30$\pm$0.005) and (2.65$\pm$0.01, 1.25$\pm$0.006). The fittings
are plotted in Figure 1. The result is rather satisfactory especially for the small
$k$ regime. It is in this regime that the $q$'s are considerably different for the
three networks. For the large $k$ regime, the three $q$'s are roughly the same.

A question naturally arises as to whether one can fit the two regime degree
distributions at the same time in one step instead of doing it separately for the
two regimes as shown above. A possible method for this global fitting was proposed
in \cite{TsallisGeorgeMendes,Borges}. Four parameters (instead of two for each
fitting) were introduced through a mathematical consideration leading to the
following differential equation of probability :
\begin{equation}\label{WholeDistributionAirNetwork}
\frac {d p(k)}{dk}=-\mu_r p^r(k)-(\lambda_q-\mu_r)p^q(k),
\end{equation}
\noindent with $r \le q$. Here $\mu_r$, $\lambda_q$, $q$ and $r$ are parameters. The
solution of the above equation is
\begin{equation}\label{IntegralFitting}
k=\int_{p(k)}^1 \frac {dx} {\mu_r x^r+(\lambda_q-\mu_r)x^q}.
\end{equation}
\noindent Further calculation using Mathematica of Eq.
(\ref{IntegralFitting}) leads to
\begin{eqnarray}
\label{HypergeomFit} \nonumber
  k&=& \frac{1}{\mu_r} \{ \frac{p^{1-r}(k)-1}{r-1}-\frac{\lambda_q/\mu_r-1}{1+q-2r} \\ \nonumber
   &&\times [H(1;q-2r,q-r,(\lambda_q/\mu_r-1)) \\
   && -H(p(k);q-2r,q-r,(\lambda_q/\mu_r-1))]\},
\end{eqnarray}
\noindent where $H(x;a,b,c)=x^{1+a}F(\frac
{1+a}{b},1;\frac{1+a+b}{c};-x^bc)$, with $F$ being the
hypergeometric function.

As described in \cite{TsallisGeorgeMendes}, the values of $\mu_r$, $\lambda_q$, $q$
and $r$ are first estimated directly from the curves depicted by the original data.
Then using Eq. (\ref{IntegralFitting}) and treating the observed values of $p(k)$ as
inputs, the values of $k$ are fitted. The results are plotted in figure 2 to 4. The
values of $q$ for the three networks are of same order as $q_1$'s from the separate
fitting described above, while the values of $r$ are all 0.6 and very different from
$q_2$'s. As in the separate fitting, the $q$ value for the composite network
China+US is smaller than the $q$ value for China network and larger than the $q$ for
US network. The fittings are globally fine, but with a small deviation observed at
the ``knee'' between the two regimes: the theoretical curves do not have knees as
distinct as the observed curves. The above fitting procedures as well as other
techniques\cite{ALM} we checked for fitting these two regime degree distributions
will be described in details in another paper.

Although the above fittings are not perfect, we can say that Tsallis distribution
function can, with the actually used mathematics, offer a satisfactory description
of the observed laws. One of the consequences of these calculations is: systems
obeying NS and composed of different $q$ subsystems all obeying NS exist. As a
matter of fact, this existence is evident by a simple reasoning. Suppose there is a
unique $q$ in the Universe, then it must be $q=1$ because we know there are systems
in the Universe that obey BGS. This contradicts the starting hypothesis of the
necessity of NS. Clearly, in general, Eq.(\ref{2}) can not exist. In what follows,
we propose an extension of it, the first one as far as we know.

\section{An alternative nonadditivity rule of entropy}
The aim of the following sections is to provide a possible formulation of NS which
allows this kind of complicated system composition. For this purpose, PJP is
abandoned as basic postulate. We take the entropy $S_q$ and the nonadditive rule of
entropy as two basic hypothesis of the theory.

Now let us consider $A$ and $B$, two nonextensive subsystems of a composite
nonextensive system $A+B$. It has been proved\cite{Abe02a} that the most general
pseudoadditivity (or composability) of entropy or energy prescribed by zeroth law is
the following :
\begin{equation}                                    \label{5}
H[Q(A+B)]=H[Q(A)]+H[Q(B)]+\lambda_Q H[Q(A)]H[Q(B)],
\end{equation}
which is in fact a very weak condition where $H[Q]$ is just certain differentiable
function satisfying $H[0]=0$, $\lambda_Q$ is a constant, and $Q$ is either entropy
$S$ or internal energy $U$\cite{Wang02a}. As shown in \cite{Wang02a}, for a given
relationship $Q=f(N)$ where $N$ is the (additive) number of elements of the system,
the finding of $H(Q)$ is trivial. Eq.(\ref{5}) has been
established\cite{Abe02a,Wang02a} for the class of systems containing subsystems
having same $q$. The generalization of it to the systems whose subsystems have
different $q$'s is straightforward if we replace the Eq.(1) of reference
\cite{Abe02a}, i.e., $S(A+B)=f\{S(A),S(B)\}$ for uniform $q$, by
$H_{q}[S_q(A+B)]=f\{H_{q_A}[S_{q_A}(A)],H_{q_B}[S_{q_B}(B)]\}$ (or by
$H_{q}[S_q(A+B)]=H_{q_A}[S_{q_A}(A)]+H_{q_B}[S_{q_B}(B)]
+g\{H_{q_A}[S_{q_A}(A)],H_{q_B}[S_{q_B}(B)]\}$) where $H_q(S_q)$ is a functional
depending on $q$'s in the same way for the composite system as for the subsystems,
where $q$, $q_A$ and $q_B$ are the parameters of the composite system $A+B$, the
subsystems $A$ and $B$, respectively. The function $f$ (or $g$) is to be determined
by the consideration of the zeroth law. Now repeating the mathematical treatments
described in the references \cite{Abe02a,Wang02a}, we find
\begin{equation}                                    \label{5x}
H_q[Q(A+B)]=H_{q_A}[Q(A)]+H_{q_B}[Q(B)]+\lambda_Q H_{q_A}[Q(A)]H_{q_B}[Q(B)].
\end{equation}
Eq.(\ref{5}) turns out to be a special case of Eq.(\ref{5x}) for same $q$ systems. In
view of the form of Tsallis entropy, a possible choice is
$H_q(S_q)=\sum_{i=1}^wp_i^q-1=(1-q)S_q$ in Eq.(\ref{5x}) as proposed in
\cite{Wang04a}. This implies :
\begin{eqnarray}                                    \label{2a}
(1-q)S_q(A+B) &=& (1-q_{A})S_{q_{A}}(A)+(1-q_{B})S_{q_{B}}(B) \\ \nonumber &+&
\lambda_S(1-q_{A})(1-q_{B})S_{q_{A}}(A)S_{q_{B}}(B)
\end{eqnarray}
Note that Eq.(\ref{2a}) is not derived from Eq.(\ref{5}) but only postulated as a
possible pseudoadditivity of Tsallis entropy. As stated in reference \cite{Tsal04},
{\it Unless the (probability) composition law is specified, the question whether an
entropy is or is not extensive has no sense.} Here we apply this reasoning
inversely, i.e., we specify a nonadditivity rule of entropy and look for the
corresponding probability composition rule. Let $\lambda_S=1$, Eq.(\ref{2a})
implies\cite{Wang04a} :
\begin{eqnarray}                                    \label{6}
p_{ij}^q(A+B)=p_i^{q_A}(A)p_i^{q_B}(B)
\end{eqnarray}
which can be called the extended factorization of joint probability for the systems of
different $q$'s. Eq.(\ref{6}) can be $formally$ written as the usual PJP
$p_{ij}(A+B)=p_{i}(A)p_{j}(B)$ if and only if $q_A=q_B$. This extended PJP is nothing
but the consequence of a kind of dependence of the subsystems. It is only one
composition law of probability among many other possible ones corresponding to
different additivity and nonadditivity of entropy, as indicated in \cite{Tsal04}. The
physical or effective probability is now $p_i^q$ instead of $p_i$. This interpretation
may help to understand why $p_i^q$ or the escort probability
$p_i^q/\sum_ip_i^q$\cite{Tsal98} should be used for defining expectation in the same
way for both the composite systems and the subsystems.

We would like to emphasize that Eq.(\ref{6}) is only a result of the postulated
nonadditivity rule Eq.(\ref{2a}). The proposition of this rule is in fact inspired
by a study of nonequilibrium systems evolving in hierarchically heterogeneous phase
space\cite{Wang04}. It has been shown that the above chosen functional form
$H_q(S_q)=\sum_{i=1}^wp_i^q-1$ is a measure of the variation of information
(dynamical uncertainty) during the evolution if the normal rule $\sum_{i=1}^wp_i=1$
holds, and that $q$ is a ratio between the Hausdorff dimension and the topological
dimension of the phase space if the latter is fractal. So the choice of
Eq.(\ref{2a}) is relevant at least in this case where Tsallis formula can be used to
measure entropy change in time.

\section{Determination of the composite $q$}
Eq.(\ref{6}) allows one to determine uniquely the parameter $q$ for the composite
system if $q_A$, $q_B$, $p_i(A)$ and $p_j(B)$ are given. By the normalization of the
joint probability, we obtain the following relationship :
\begin{eqnarray}                                    \label{7}
\sum_{i=1}^{w_A}p_i^{q_A/q}(A)\sum_{j=1}^{w_B}p_j^{q_B/q}(B)=1
\end{eqnarray}
which means $q_A<q<q_B$ if $q_A<q_B$ and $q=q_A=q_B$ if $q_A=q_B$. In this way, for
a composite system containing $N$ subsystems (k=1,2,...,N) having different $q_k$,
the parameter $q$ is determined by
\begin{eqnarray}                                    \label{7a}
\prod_{k=1}^N\sum_{i_k=1}^{w_k}p_{i_k}^{q_k/q}=1,
\end{eqnarray}
from which we can say that $mini(q_k)<q<maxi(q_k)$. In order to see more clearly the
method, we suppose equiprobability for each subsystems, i.e.,
$p_{i_k}=(\frac{1}{w_k})^{1/q_k}$. In this case, we have
\begin{eqnarray}                                    \label{7b}
\prod_{k=1}^N\sum_{i_k=1}^{w_k}\left(\frac{1}{w_k}\right)^{q_k/q}=1,
\end{eqnarray}
which means
\begin{eqnarray}                                    \label{7c}
\prod_{k=1}^N\left(\frac{1}{w_k}\right)^{q_k}=\prod_{k=1}^N\left(\frac{1}{w_k}\right)^{q}
\end{eqnarray}
or
\begin{eqnarray}                                    \label{7d}
q=\sum_{k=1}^Nq_k\ln{w_k}/\sum_{k=1}^N\ln{w_k}.
\end{eqnarray}
$q$ is a kind of barycenter of the sub $q$ values. If $w_1=w_2= ... =w_N$,
Eq.(\ref{7d}) becomes
\begin{eqnarray}                                    \label{7e}
q=\frac{1}{N}\sum_{k=1}^Nq_k.
\end{eqnarray}

\section{Temperature definition through zeroth law}
A zeroth law for different $q$ systems has been briefly discussed in \cite{Wang04a}
with the help of a deformed entropy and energy. Following is a recapitulation of that
method using directly the entropy $S_q$ and energy $U_q$. If $(A+B)$ is at equilibrium
or stationary state optimizing Tsallis entropy $dS_q(A+B)=0$, considering
Eq.(\ref{2a}), we get :
\begin{eqnarray}                                    \label{8}
\frac{(1-q_{A})dS_{q_{A}}(A)}{1+(1-q_{A})S_{q_{A}}(A)}
+\frac{(1-q_{B})dS_{q_{B}}(B)}{1+(1-q_{B})S_{q_{B}}(B)}=0
\end{eqnarray}
Notice the difference between this relation and
$\frac{dS_q(A)}{1+(1-q)S_q(A)}+\frac{dS_q(B)}{1+(1-q)S_q(B)}=0$ which has been used
for the systems of same $q$.

In order to find the suitable nonadditive rule of energy, we consider Eq.(\ref{6})
and the relationship $\sum_ip_i^q=Z_q^{1-q}+(1-q)\beta U_q$ calculated from the
distribution Eq.(\ref{4}) and the unnormalized expectation $U_q$, we get
\begin{eqnarray}                                    \label{xx10}
& Z_q^{1-q}(A+B)+(1-q)\beta(A+B) U_q(A+B) \\\nonumber & =
[Z_{q_A}^{1-q_A}(A)+(1-q_A)\beta(A) U_{q_A}(A)][Z_{q_B}^{1-q_B}(B)+(1-q_B)\beta(B)
U_{q_B}(B)].
\end{eqnarray}
Then the total energy conservation $dU_q(A+B)=0$ leads to
\begin{eqnarray}                                    \label{xx11}
\frac{(1-q_{A})\beta(A)dU_{q_A}(A)}{\sum_ip_i^{q_A}(A)}
+\frac{(1-q_{B})\beta(B)dU_{q_B}(B)}{\sum_ip_i^{q_B}(B)}=0
\end{eqnarray}
which suggests following energy nonadditivity
\begin{eqnarray}                                    \label{xx12}
\frac{(1-q_{A})dU_{q_A}(A)}{\sum_ip_i^{q_A}(A)}
+\frac{(1-q_{B})dU_{q_B}(B)}{\sum_ip_i^{q_B}(B)}=0
\end{eqnarray}
as the analog of the additive energy $dU(A)+dU(B)=0$ of BGS. From Eq.(\ref{xx12})
and Eq.(\ref{8}) follows
\begin{eqnarray}                                    \label{xx12x}
\frac{\partial S_{q_A}(A)}{\partial U_{q_A}(A)}=\frac{\partial S_{q_B}(B)}{\partial
U_{q_B}(B)}
\end{eqnarray}
so the inverse temperature can be defined by $\beta=\frac{\partial S_{q}}{\partial
U_{q}}$ for any system with whatever $q$.

We would like to emphasize here that Eq.(\ref{xx12x}) allows us to measure the
temperature of a system with a thermometer having different $q$ values, as far as the
system and the thermometer are in thermal equilibrium or local thermal equilibrium at
the point of contact. This result mathematically supports a previous numerical work of
\cite{Tsal03,Moyano} showing the possibility to measure the temperature of a non
Boltzmannian nonequilibrium system with a Boltzmannian thermometer.

It should be noticed that the present result is obtained with the unnormalized
expectation of energy. In view of the important role of the escort probability in
NS, an extension of NS to different $q$ system with the expectation defined with
escort probability is indeed necessary. This possibility will be investigated in our
future work on the basis of the extended PJP.

\section{Concluding remarks}
In the first part of this work, we fitted the power law degree distribution in two
regimes of some airport networks with the $q$-exponential distribution of
nonextensive statistics. The results prove with sufficient exactitude the existence
of different $q$ subnetworks and composite networks all obeying $q$-exponential
distribution. This situation requires formulation of NS allowing the composition of
different $q$ systems. A crucial step is to show that temperature can be uniform in
the nonextensive systems at equilibrium or local equilibrium states optimizing
Tsallis entropy, independently of whether or not the systems contain subsystems with
different $q$ values.

This formulation is constructed on the basis of an extended nonadditivity rule of
entropy taken as a basic hypothesis of the theory. In this case, the conventional
PJP is replaced by an extended PJP which signifies a kind of dependence between
subsystems. Even when PJP is recovered in the special case of unique $q$, it has
nothing to do with the independence of subsystems in the context of nonextensive and
nonadditive systems. This is an idea expressed in a previous discussion
\cite{Tsal04}. As indicated above, this is only one of the possible formulations of
NS with a pseudoadditivity rule. Further investigation is necessary to see the
physics behind each possible composition law of entropy and probability.

We would like to mention here that the extended PJP and the usual normalization
condition together prescribe two possible ways for generalizing NS to different $q$
systems: the first is with the unnormalized expectation as has be done in the present
work; the second is with the normalized expectation given by the escort probability.
These two definitions of expectation allow simple composition rules of expectation of
physical quantities. The possibility of generalizing NS with the escort probability is
actually under investigation.

\section*{Acknowledgement}
We thank Constantino Tsallis for valuable discussion and suggestions, and Ernesto P.
Borges for fruitful participation and assistance in the computation. We would like to
thank the referees for constructive comments and suggestions. This work is supported
in part by the R\'egion des Pays de la Loire of France under Grant $N^o$ 04-0472-0 and
National Natural Science Foundation of China.

\begin{figure}[hb] \label{f1}
\includegraphics[width=6cm,height=15cm]{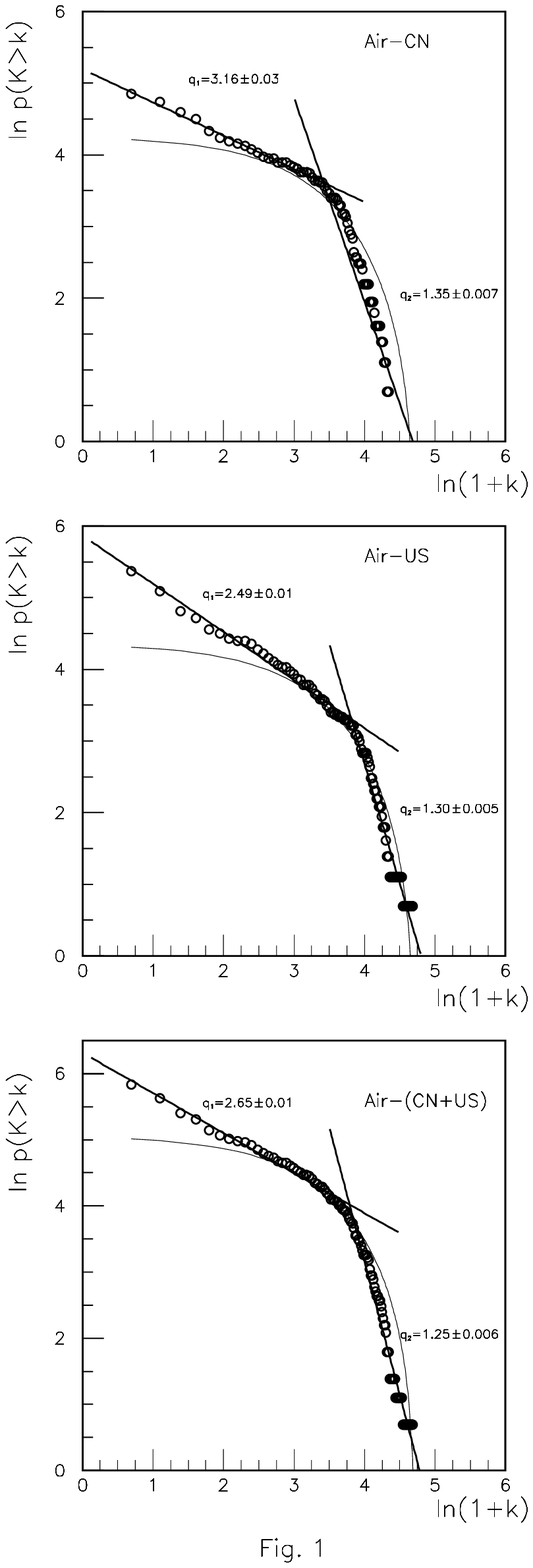}
\caption{Degree distributions (circles) of China airline network (top panel), US
airline network (middle panel), and airline network of China plus US (bottom panel).
The straight lines are fittings of the observed laws with the q-distribution given by
Eq.(5). In order to show the deviation of the observed two regime distribution from
exponential law, this latter is represented in the figure by curved lines. }
\end{figure}

\begin{figure}[hb] \label{f2}
\includegraphics[width=16cm,height=14cm]{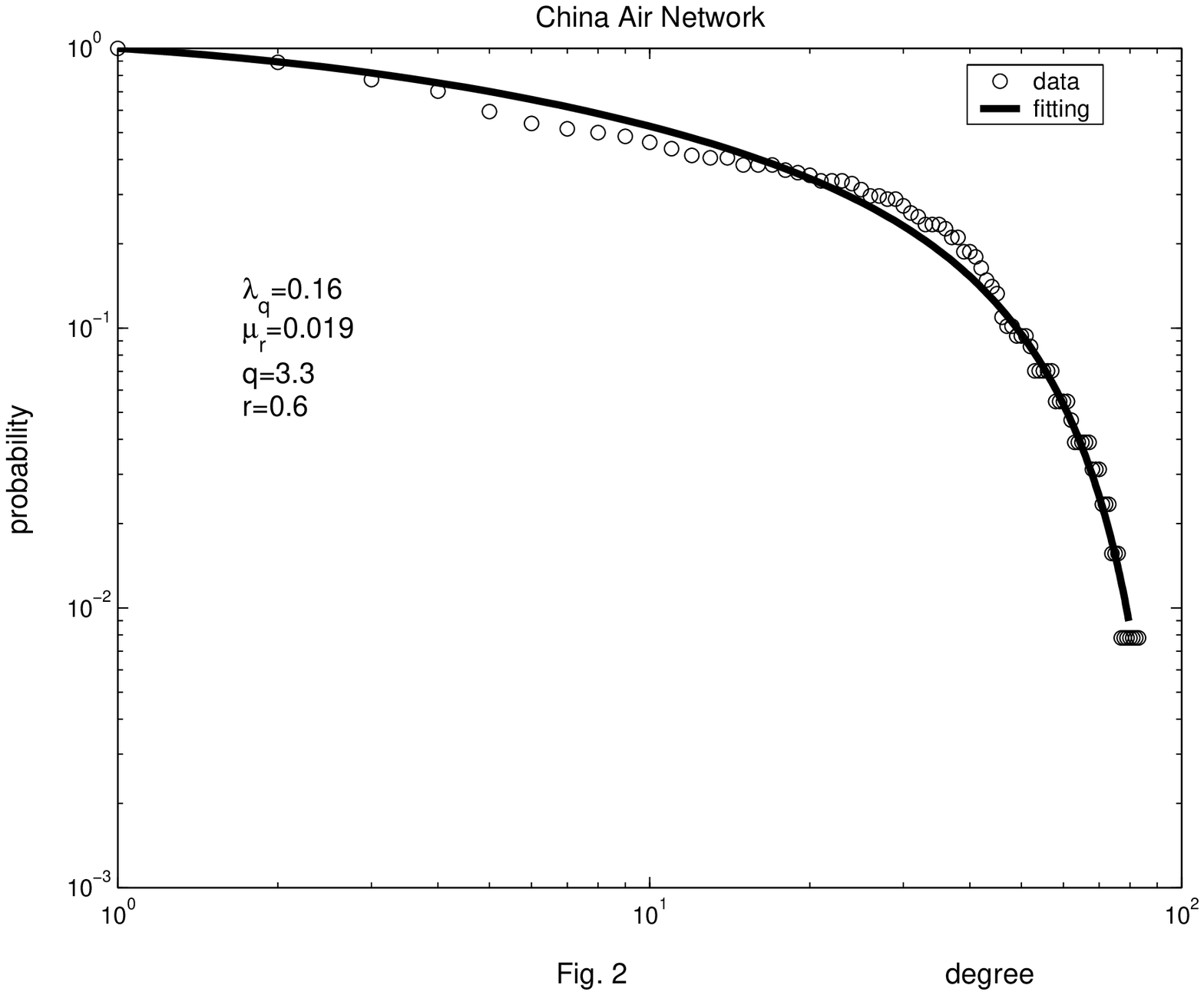}
\caption{Degree distribution (points) of China air network. The line comes from the
fitting using Eq. (\ref{IntegralFitting}) where the four parameters $\mu_r$,
$\lambda_q$, $q$ and $r$ were estimated directly from the data points.}
\end{figure}

\begin{figure}[hb] \label{f3}
\includegraphics[width=16cm,height=14cm]{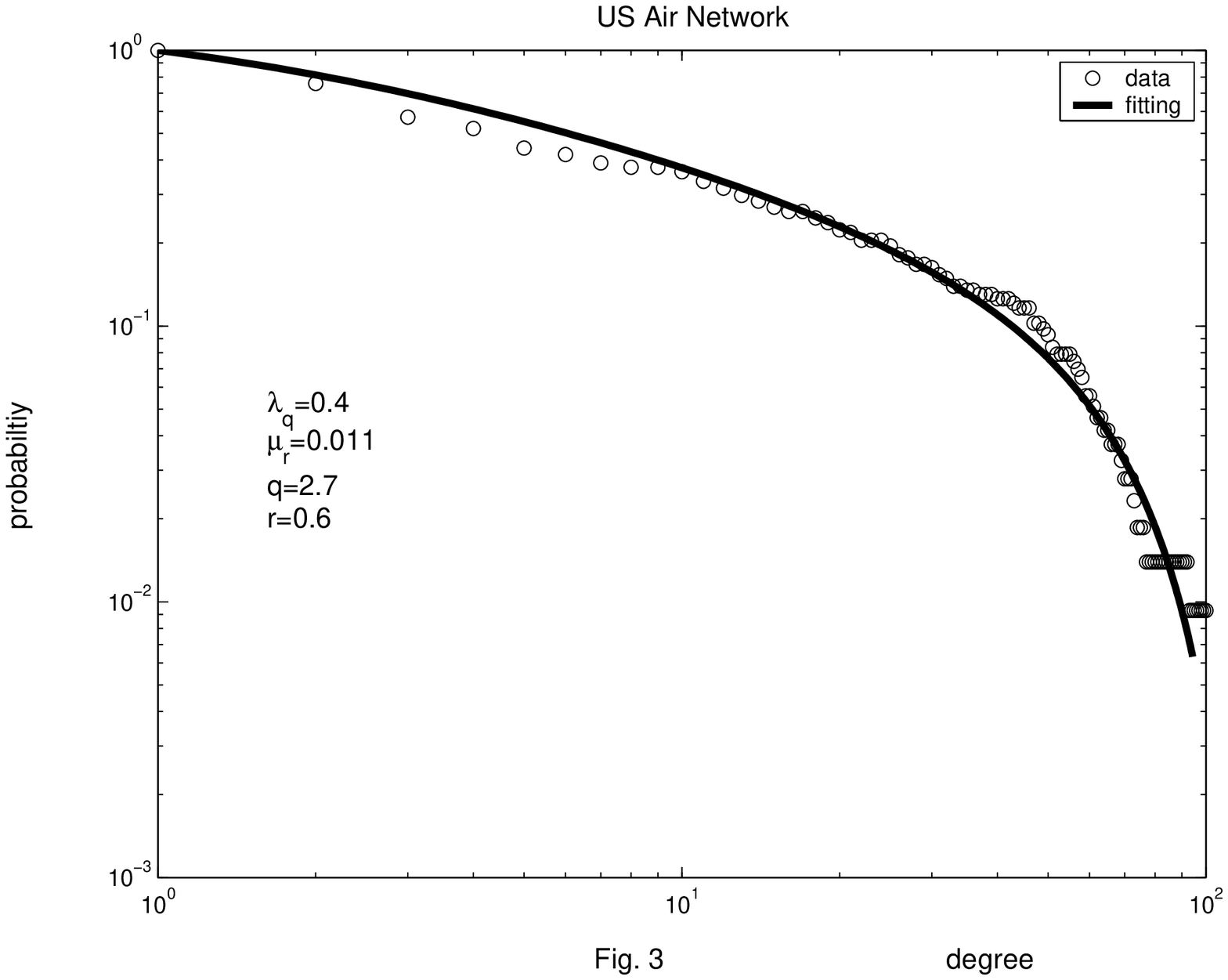}
\caption{Degree distribution (points) of US air network. The line comes from the
fitting using Eq. (\ref{IntegralFitting}) where the four parameters $\mu_r$,
$\lambda_q$, $q$ and $r$ were estimated directly from the data points.}
\end{figure}

\begin{figure}[hb] \label{f4}
\includegraphics[width=16cm,height=14cm]{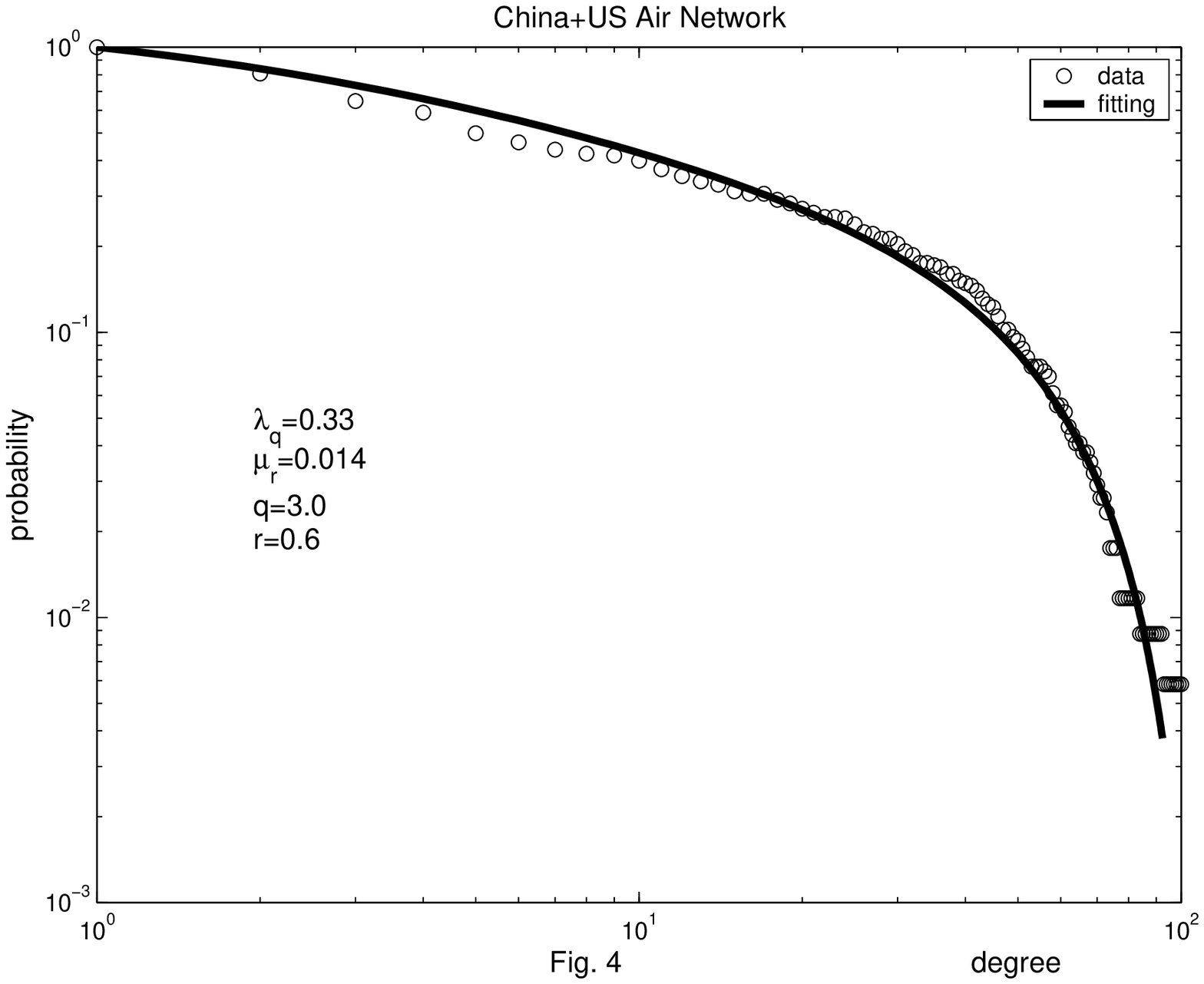}
\caption{Degree distribution (points) of China+US air network. The line comes from
the fitting using Eq. (\ref{IntegralFitting}) where the four parameters $\mu_r$,
$\lambda_q$, $q$ and $r$ were estimated directly from the data points.}
\end{figure}

\end{document}